\documentclass[useAMS,usenatbib]{mn2e}
\usepackage{graphicx}

\def\simlt{\lower.5ex\hbox{\simlt}}
\def\gtsima{$\; \buildrel > \over \sim \;$}
\def\simgt{\lower.5ex\hbox{\gtsima}}

\title[A low surface brightness halo surrounding NGC~5694]{A low surface brightness halo surrounding the globular cluster NGC~5694\thanks{Based on observations made with ESO Telescopes at the Paranal Observatory under programme ID 081.B-0428(A)}}
\author[M. Correnti et al.]{M. Correnti$^{1,2}$\thanks{E-mail:
correnti@iasfbo.inaf.it}, M. Bellazzini$^{1}$, E. Dalessandro$^{3}$, A. Mucciarelli$^{3}$, 
\and L. Monaco$^{4}$, M. Catelan$^{5}$\\ 
$^{1}$INAF-Osservatorio Astronomico di Bologna, Via Ranzani 1, 40127, Bologna, Italy.\\
$^{2}$INAF-Istituto di Astrofisica Spaziale e Fisica Cosmica di Bologna, Via Gobetti 101, 40129, Bologna, Italy.\\
$^{3}$Dip. di Astronomia - Univ. di Bologna, Via Ranzani 1, 40127, Bologna, Italy.\\
$^{4}$European Southern Observatory, Casilla 19001, Santiago, Chile.\\
$^{5}$Pontificia Univ. Cat\'{o}lica de Chile, Departamento de Astronom\'{i}a y Astrof\'{i}sica, Av. Vicu\~na Mackenna 4860, 782-0436 Macul, Santiago, Chile.
}
\begin{document}

\date{Accepted Received ; in original form }

\pagerange{\pageref{firstpage}--\pageref{lastpage}} \pubyear{2011}

\maketitle

\label{firstpage}

\begin{abstract}
We report on the discovery of an extended stellar halo surrounding the distant
Galactic globular cluster NGC~5694, based on new deep ($V\simeq 24.5$) wide-field
($24\arcmin \times 20\arcmin$)  photometry acquired with VIMOS at VLT. Stars
with colour and magnitude consistent with the Main Sequence of the cluster are
clearly identified out to $r\simeq9\arcmin(\simeq93$~pc) from the cluster
center, much beyond the tidal radius of the King model that best fits the inner
profile ($r_t=3.15\arcmin$). We do not find a clear end of the structure within
our field.  The overall observed profile cannot be properly fitted with either a
King (1966) model, an Elson et al. (1987) model, or a Wilson (1975) model; however it is very smooth and
does not show any sign of the break near the tidal radius that is typically
observed in stellar systems with tidal tails. The density map we derived does not
show evidence of tidal tails, within the considered field. The extra-tidal
component contains $\simeq 3.5$\% of the cluster light (mass) and has a surface
density profile falling as $\sim r^{-3.2}$.  The possible origin of the detected
structure is discussed, as a clear-cut conclusion cannot be reached with the
available data.
\end{abstract}

\begin{keywords}
{\em (Galaxy:)} globular clusters: individual: NGC~5694 --- Galaxy:formation
\end{keywords}

\section{Introduction}

NGC~5694 is a relatively poorly studied globular cluster (GC) residing in the
outer Halo of the Galaxy. Based on its large distance ($\ga 30$~kpc) and radial
velocity ($V_r=-144$~km~s$^{-1}$), \citet{hh76} suggested that the cluster is on
a hyperbolic orbit not bound to the Milky Way (MW). \citet[][OG90
hereafter]{og90} were the first to publish CCD photometry of the cluster (over
a  $\sim 4\arcmin \times 2.5\arcmin$ field of view). They concluded that
NGC~5694 is a metal-poor ([Fe/H]=$-1.65$) stellar system with an age comparable
to that of the oldest GCs of similar metallicity  \citep[age$\ga 12.5$~Gyr,
see][]{dotter}, lying at a distance $D\simeq 32$~kpc. From low-resolution Ca
triplet spectroscopy of nine cluster giants, \citet{geisler} obtained
[Fe/H]=$-1.87\pm 0.07$, and slightly revised the systemic velocity,
$V_r=-140.7\pm 2.4$~km~s$^{-1}$. The low metallicity and old age were confirmed
by \citet{dea}, based on HST/WFPC2 photometry of the cluster center. The
Horizontal Branch (HB) is significantly populated only at colours bluer than the
RR Lyrae instability strip and, indeed, no such variable was found in the
cluster \citep{haz}.\\  
Eventually, \citet[][hereafter L06]{l06} obtained high-resolution spectroscopy
of a star on the Red Giant Branch (RGB) of NGC~5694. They were able to derive
the abundance of several elements, including Fe ($\rm [Fe/H]=-1.93\pm 0.07$) and
other iron-peak elements, $\alpha$-elements, $r-$ and $s-$ elements. They found 
an abundance of $\alpha$-elements $\rm (Ca+Ti)/(2Fe)\simeq 0.0$, i.e. much lower
than that typical of old GCs in the MW halo ($\sim +0.3$) and similar to stars
of similar metallicity in nearby dwarf spheroidal (dSph) galaxies, in globular
clusters of the LMC and Fornax dSph, and in a handful of MW GCs
\citep[see for example][]{cohen04,venn04,monaco05,sbord07}, that are known (or strongly suspected) to have been
accreted from disrupted/disrupting dwarf Galactic satellites \citep[see][for
references and discussion]{pritzl,lm10,kirby}.  Moreover, NGC~5694 appears to
have [Cu/Fe] and [Ba/Eu] ratios among the lowest of any stellar system for which
the abundances of these elements have been measured. L06 interpret these
chemical anomalies and the high velocity as evidence for an extra-Galactic
origin of the cluster. They did not found any significant correlation with
existing satellites and/or other GCs, thus concluding that the original parent
galaxy is presently dissolved. Indeed, the association of several GCs in the
outer halo of the MW and M31 with relics of partially or totally disrupted dwarf
galaxies they originally belonged to is now quite firmly established
\citep{bfi,perina,lm10,mackey}.\\ Triggered by this fascinating hypothesis on
the origin of NGC~5694, we obtained new deep photometry over a wide Field of
View (FoV), to search the surroundings of the clusters for stellar features that
might reveal hints on the origin and evolution of the system. In this paper
we report on the detection of stars far beyond the reported tidal radius of the
cluster, that appears to have a profile smoothly extending at least out to
$\simeq 9\arcmin$ from the center.

\section{Observations and Data Reductions}

Observations were performed on the night of July 29, 2008 with the VIMOS camera
mounted at ESO-VLT UT3, Paranal Observatory (Chile), under stable and clear
conditions (seeing $\simeq 1\arcsec$~FWHM). VIMOS is a mosaic of four $\simeq
2000 \times 2300$~px$^2$ CCDs (with pixel scale $0.205\arcsec/{\rm px}$), each
covering a FoV of  $7\arcmin \times 8\arcmin$, separated by $2\arcmin$ gaps. We
obtained one B and one V exposure  per pointing, in four different pointings,
thus covering a total FoV of $\sim 24\arcmin \times 20\arcmin$ centered on the
cluster, without gaps. The exposure time was always set to $t_{exp}= 200$~s.\\
The images were corrected for bias and flat-field with standard IRAF procedures.
Photometry of individual stars was performed with the well-known Point
Spread Function fitting code DAOPHOT \citep{s87}. The images were searched for
sources with intensity peaks at $\ge 3~\sigma$ above the background. The pixel
coordinates of the catalogues with instrumental B,V magnitudes for each chip
were transformed into J2000 Equatorial coordinates by means of third-order
polynomials obtained using several ($\ga 60$ per chip) astrometric
standards\footnote{Using the CataPack suite of codes, developed by P.
Montegriffo at INAF-Osservatorio Astronomico di Bologna.} from the GSC2.2
catalogue\footnote{\tt http://tdc-www.harvard.edu/catalogs/gsc2.html}. 
As an independent test we found that the standard deviation of the residuals in both RA and Dec for the 513 stars in common with the 2MASS catalogue \citep{2mass} over the whole FoV of the mosaic is $\le 0.2\arcsec$.
The instrumental magnitudes of all the catalogues were reported to the same
instrumental system using stars in common in the overlapping regions between
different chips. Then all the individual catalogues were merged together into a
total catalogue, averaging the magnitudes of the stars measured in more than one
chip. Finally, the instrumental magnitudes were transformed into the
Johnson-Kron-Cousins system using 74 standard stars from the set by \citet{s00},
that were in common with the total catalogue. The agreement in the photometric
zero points with the independent photometry by OG90 is excellent, with
average differences smaller than 0.01 mag and rms scatter $=0.035$ mag in both B and V (from 166 stars in common with $V\le21.0$).\\ 
We estimated the reddening by interpolating on the \citet{sfd} maps on a grid
covering the whole FoV at the resolution of the maps ($\simeq 5\arcmin$).
According to this source, the reddening is very uniform over the field, with
maximum differences smaller than $0.015$ mag. The average value is
$E(B-V)=0.099$ which is in excellent agreement with the result by OG90 and is
adopted in the following. To estimate the distance we
matched, on the Colour-Magnitude Diagram (CMD), the HB of NGC~5694 with that of
M~68 \cite[from][]{m68}, a GC with a similar metallicity and a well populated BHB. The
best match is obtained assuming $(m-M)_0=17.75 \pm 0.1$\footnote{E(B-V)=0.04 and
$(m-M)_0=15.14$ were adopted for M~68, according to \citet[][their
Tab.~2]{f99}.}, corresponding to $D=35.5$~kpc, in agreement with the distance
reported in the \citet[][]{h96} catalogue (2010 edition), and slightly larger
than what was found by OG90. At this distance, 1 arcmin corresponds to 10.3 pc.
In the following we will adopt the coordinates of the cluster center provided by
\citet[][NG06 hereafter]{ng06}. Since the incompleteness in the innermost
$0.5\arcmin$ is quite large, due to the high degree of crowding, in the present
work we will limit our analysis only to the stars at distances larger than
$1\arcmin$ from the cluster center.

\begin{table}
\label{tabsb}
 \centering
 \begin{minipage}{70mm}
  \caption{Surface Brightness profile from star counts, for $r\ge 1.5\arcmin$.}
  \begin{tabular}{@{}lcc@{}}
  \hline
   log r     &  $\mu_V$\footnote{Background subtracted. Not corrected for reddening.} & err($\mu_V$)\\
    $[$arcmin$]$ & $[$mag/arcmin$^2]$  &  $[$mag/arcmin$^2]$ \\
  \hline
 0.1761 & 23.92 &  0.05 \\
 0.2304 & 24.27 &  0.06 \\
 0.2788 & 24.65 &  0.06 \\
 0.3222 & 24.93 &  0.07 \\
 0.3617 & 25.06 &  0.07 \\
 0.3979 & 25.36 &  0.08 \\
 0.4314 & 25.70 &  0.09 \\
 0.4624 & 25.89 &  0.09 \\
 0.4914 & 26.22 &  0.10 \\
 0.5185 & 26.52 &  0.12 \\
 0.5441 & 26.61 &  0.12 \\
 0.5911 & 26.94 &  0.13 \\
 0.6532 & 27.52 &  0.07 \\
 0.7404 & 28.46 &  0.11 \\
 0.8129 & 29.17 &  0.16 \\
 0.8751 & 29.57 &  0.20 \\
 0.9294 & 29.91 &  0.23 \\
\hline
\end{tabular}
\end{minipage}
\end{table}

\section{Results}

\begin{figure*}
\includegraphics[width=128mm]{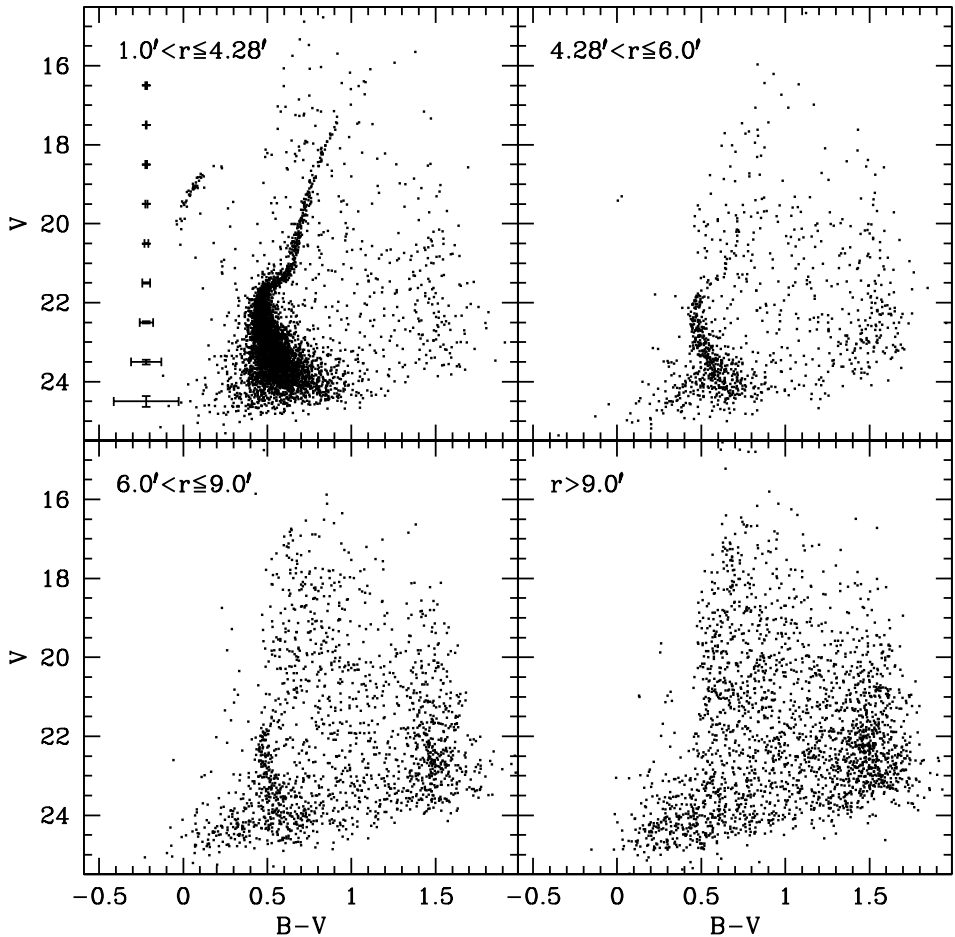}
  \caption{Colour-Magnitude Diagram of NGC~5694 in different radial annuli.
$r=4.28\arcmin$ corresponds to the tidal radius derived by TKD. The average
errors are reported as errobars on the blue side of the upper-left panel. A
comparison with the predictions of the TRILEGAL Galactic model \citep{trilegal}
for this direction suggests that the sparse vertical plume of blue stars at
$B-V\sim 0.3$ and $18.0\la V\la 23.0$ that is seen in the lower two panels is
likely due to nearby White Dwarfs belonging to the Galactic Disc.}
  \label{cms}
\end{figure*}

In Fig.~\ref{cms} we show the derived CMD for different radial annuli around the
cluster center. This is the deepest and cleanest CMD ever obtained for NGC~5694,
for such a wide FoV.  The upper-left panel shows the most central annulus
considered here. In this diagram, the most prominent feature is the cluster Main
Sequence [MS, from (V, B-V)$\sim (22.0, 0.45)$ to $\sim (24.5, 0.65)$], with the
short Sub Giant Branch just above. A narrow and steep RGB and a Blue HB are also
clearly visible. Most of the remaining stars are attributable to Galactic
foreground, except for the blob of faint blue sources at $V\ga 23.5$ and $B-V\la
0.5$ that are more likely background galaxies. The overall CMD  is fully
consistent with the age and metallicity reported in the literature. The limit at
$r=4.28\arcmin$ corresponds to the tidal radius \cite[$r_t$, see][]{k66} derived
by \citet[][TKD, hereafter]{tkd}. The upper-right and lower-left panels of
Fig.~\ref{cms} show that cluster stars (especially MS and SGB) are clearly
present in significant numbers far beyond the TKD tidal radius, at least
out to $r\simeq 9\arcmin$, corresponding to $\simeq 93$~pc. In fact, the
comparison between the two lower panels suggests that cluster MS stars may be
present also beyond that radius. With the available data we  were unable to
detect any clear sign of an end of the distribution of cluster stars (see below). Hence, we
cannot exclude that it extends beyond the $24\arcmin \times 20\arcmin$ field
considered here.

\begin{figure}
\includegraphics[width=80mm]{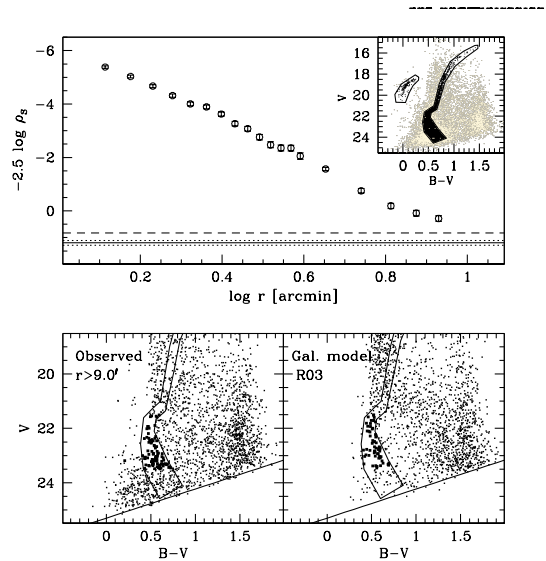}
 \caption{{\em Upper panel:} surface density profile from star counts (not background-subtracted). The background level, estimated in the region $r>9.0\arcmin$, is marked by a continuous line; the dotted lines are at $\pm 1\sigma$ from the background density. The dashed line is located at 5 $\sigma$ above the background level. In the inset on the
 upper-right corner, the selection on the CMD of the most likely cluster
 members used for star counts is displayed. 
 {\em Lower left panel:} observed CMD for $r>9.0\arcmin$. The stars enclosed in the selection contour adopted for star counts (see also Fig.~\ref{prof}) and having $21.5\le V\le 23.5$ are plotted as heavier dots. The diagonal line approximately marks the locus of the limiting V magnitude as a function of colour. {\em Lower right panel:} the same as the lower left panel for a synthetic CMD from the R03 Galactic model, for a field of view of the same area as the lower left panel.
 }
 \label{vedip}
\end{figure}

\begin{figure}
\includegraphics[width=80mm]{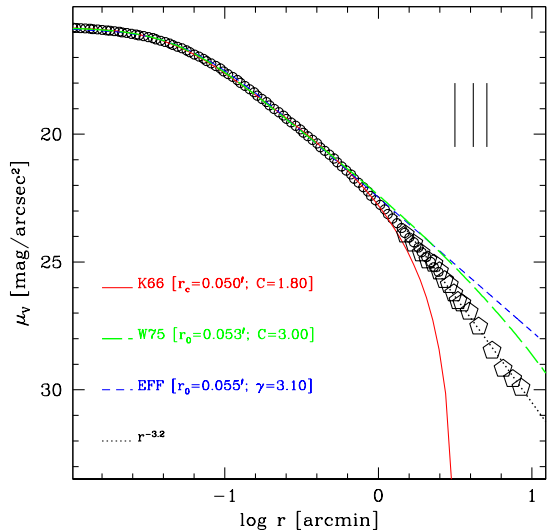}
 \caption{Surface brightness profile of NGC~5694. Open circles are the smoothed
 profile by NG06, whereas large open pentagons are obtained from star counts on
 our data; $\mu_V$ values are {\em not} corrected for reddening and the
 uncertainties are smaller than the size of the points. The red continuous
 line, the blue short-dashed line, and the green long-dashed line are the K66, EFF,  and W75  models, respectively, that
 best fit the inner part of the profile ($r\le 1\arcmin$). The dotted line is the power-law
 best-fitting the profile beyond $\sim 1.2\arcmin$. The small vertical segments
 mark the position of the tidal radius as derived with K66 model fitting, from
 left to right, in the present paper, by TKD, and by McV.  }
 \label{prof}
\end{figure}

By adopting the same procedure used and described in detail by \citet{lucky},  we obtained the surface density profile from star counts on concentric annuli  (before subtraction of the background and limited to $r\ga1.2\arcmin$) that is shown in the upper panel of Fig.~\ref{vedip}. 
The stars were selected as likely cluster members based on the selection boxes
shown in the upper-right inset and were counted out to
$r=10\arcmin$, i.e. the largest circle that is fully enclosed within our field.
The area outside this circle was used to estimate the background density. It is clear that even in the outermost point of the profile the excess of surface density above the background is very significant ($>5\sigma$). This suggests that the distribution of cluster stars likely extends beyond this limit. If true, this would imply that the adopted background level is an overestimate of the true surface density of the background. To verify this possibility we would need observations of similar deepness over a FoV wider than the one considered here, that are not available. In the lower panels of Fig.~\ref{vedip} we compare the observed CMD for the $r>10\arcmin$ region, with the corresponding synthetic CMD predicted by the Galactic model of \citet[][R03 hereafter]{r03}. To do that (a) we retrieved model predictions for a FoV of 1~deg$^2$ in the direction of NGC5694, (b) we rescaled the extinction of the model to have the same asymptotic value as the observed one, and corrected the stars magnitudes accordingly, (c) we randomly extracted a subsample by rescaling the number of stars to the ratio of the considered FoVs, (d) we cut the data according to the limiting magnitude line of the observed CMD, and, (f) we added an observational error to each magnitude of each star, randomly extracted from a Gauss distribution having the same $\sigma$ of observed stars of the same magnitude (see Fig.~\ref{cms}).
We made $10^4$ random extractions of the synthetic sample, like the one shown in the lower right panel of Fig.~\ref{vedip}. For each extraction we counted the number of stars falling within the over-plotted selection box and having $21.5\le V<23.5$. We limited the comparison to MS stars since the MS is the only cluster sequence that is populated at large radii, and we adopted $V<23.5$ to minimize the contamination by background galaxies.
The distribution of this number $N_{mod}$ has average $\langle N_{mod}\rangle =57.3$, standard deviation $\sigma=7.5$, a minimum $N_{mod}^{MIN}=36$ and a maximum 
$N_{mod}^{MAX}=86$, while the observed value is $N_{obs}=88$, i.e. in excess of the model predictions in the 100 per cent of the cases. Very similar results are obtained if the TRILEGAL Galactic model \citep{trilegal} is adopted, instead of R03. It should be noted that $N_{mod}$ is an estimate for the upper limit of the expected $N_{obs}$, since the synthetic sample is not affected by incompleteness, while the observed one is. On the other hand it is still possible  that some background galaxy spilled into the selection box, thus contaminating the counts of the observed sample\footnote{Note that this is not a problem for the profile shown in Fig.~\ref{prof}, below, since any contamination by unresolved galaxies should equally affect the cluster area and the (observed) background area, while these sources are not included in Galactic models.}. The only conclusion that can be drawn is that the predictions of the R03 and TRILEGAL models are fully consistent with the hypothesis that the distribution of NGC5494 stars extends beyond the limits sampled by the profile of Fig.~\ref{vedip}.

The Surface Brightness (SB) profile of NGC~5694 was studied by TKD based on
accurately assembled heterogeneous data from CCD and photographic images.
\citet[][McV hereafter]{mcvdm} re-analyzed the same data-set finding slightly
different solutions for the best-fitting K66 model (obtaining
$r_t=5.09\arcmin$). On the other hand, NG06 used archival HST data to obtain
accurate and homogeneous SB profiles in the innermost $2.4\arcmin$. In
Fig.~\ref{prof} we adopted the smoothed version of the observed profile provided
by NG06 as a reference to convert into absolute units the background-subtracted profile of the outer region, that we obtained from star counts. The excellent match in the overlapping region $1.0\arcmin \la r\la 2.5\arcmin$ guarantees that our profile is not affected by radial variations of the completeness, since the NG06 profile is obtained from integrated light, that, by definition, cannot suffer from incompleteness \citep[see][for a thorough discussion]{lucky}.

It is quite clear that, while the innermost $1.0\arcmin$ of the profile
(including $\ga 80$\% of the cluster light) is well fitted by a K66 model with
the reported parameters, having $r_t=3.15\arcmin$, the observed profile smoothly
extends far beyond this radius without any sign of a discontinuity. By numerical
integration of the observed profile we estimated that  $\simeq 3.5$\%  of the
cluster light is found outside this radius; we obtained also purely empirical
estimates of the half-light radius ($r_h=0.28\arcmin$) and of the integrated
absolute V magnitude ($M_V=-8.0$), both in good agreement with the values
reported by \citet{h96}. It is interesting to note that the newly derived
profile extends beyond any estimate of $r_t$ that can be found in the
literature, by a factor $\ga 2$. We tried also to fit the SB distribution
with an \citet[][EFF hereafter]{eff} profile, as these models were introduced to
describe (young) extragalactic clusters that were more extended than what is
allowed for by classical K66 models (see McV for discussion). Again, we find that
the EFF model that fits well the inner parts are unable to fit the observed
profile beyond $r\sim 1.2\arcmin$, because it predicts too much light (stars) in
this region. 

Finally, McV fitted the observed profiles of all the GCs they considered, also with isotropic and spherical \citet[][W75 hereafter]{w75} models,
finding that ``{\em ...in the majority of cases\footnote{Including NGC5694, where the three models have remarkably similar performances in fitting the observed profile, in the radial range covered by the data used by McV ($r\la 3\arcmin$; see their Tab.~10 and Fig.~12l).} the Wilson models which are spatially more extended than King models but still include a finite, ``tidal'' cutoff in density, fit clusters of any age ... as well as or better than King models}''. The long-dashed curve in Fig.~\ref{prof} is the profile of a spherical and isotropic W75 model that provides an excellent fit to the observed profile within $r\le 1.0\arcmin$, but it fails to match the outer branch, predicting a significant excess of light for $r>1.2\arcmin$, even if less pronounced than the EFF model.
By linear regression we found that the power-law that best fits the
outer profile ($r\ga 1.2\arcmin$) corresponds to $I \propto r^{-3.2}$, where
$\mu_V\propto -2.5$ $logI$.

\section{Summary and Discussion}

The low SB extended component in NGC~5694 might be, of course, of tidal origin.
In fact, it is well known that GCs lose stars because of  tidal stripping 
\citep[see][and references therein]{gnedin,c99,mont}, and, indeed, extended 
extra-tidal components or tidal tails have been observed in several cases 
\citep[][among others]{grill95,leon,lucky,pal5,bel5466,chun,jg,pal14}.\\
The main arguments supporting this interpretation are (a) the CMD of the stars
found beyond the cluster tidal radius do not appear to differ from the CMD of
stars residing in the main body of the cluster, and (b) the surface density 
steadily (and steeply) decreases from the cluster center out to the farthest
radius sampled by our data, i.e. the cluster appears as the center of symmetry
for the distribution of extra-tidal stars.\\

\begin{figure}
\includegraphics[width=80mm]{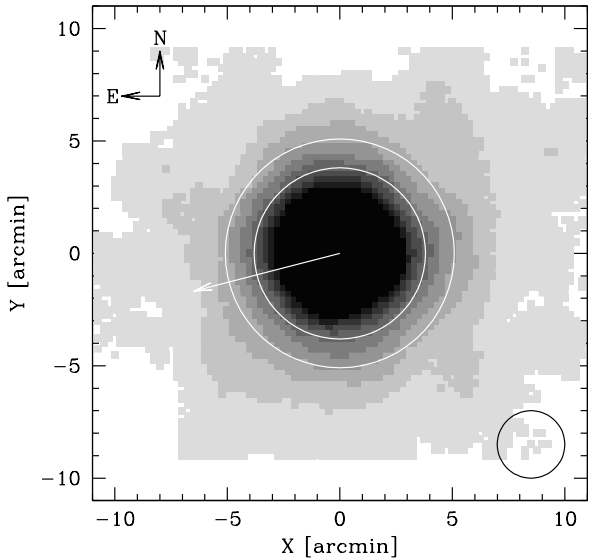}
 \caption{Density map of cluster MS stars with $V\le 21.0$. The density is
computed on a regular grid with step=$0.25\arcmin$ over a circle of radius
$1.5\arcmin$ (plotted in the lower right corner, for reference). The density
scale goes from $3\sigma$ to $21\sigma$ with steps of $2\sigma$, from lighter to
darker tones of grey. The white circles have radii equal to the cluster tidal
radius as derived by TKD (inner) and by McV (outer). The white arrow indicates
the direction toward the center of the Galaxy. The X,Y coordinates
are defined as in Eq.~1 of \citet{ven}. In a recent theoretical investigation,
\citet{mont} found that tidal tails within a few tidal radii from the cluster
center should align with the direction toward the galactic center.
 }
 \label{map}
\end{figure}


However, there are also relevant differences with respect to the ordinary cases
of clusters tidal tails studied in the literature and NGC5694. First of all, the observed
extra-tidal portions of the profiles tipically display a much shallower decline
with radius, $\sim r^{-1}$, instead of $\sim r^{-3}$ as found here \citep[][and
references therein]{grill95,leon,chun}\footnote{However, in their theoretical
analysis, \citet{c99} predicted a $\sim r^{-3}$ decline of the surface density
of extra-tidal components in the immediate surroundings of the tidal radius.}. 
Even more remarkably, we do not detect any break in the SB profile in the
vicinity of the tidal radius, which is instead observed 
\citep{grill95,leon,chun} and predicted \citep{c99,katy} as a general feature of
extra-tidal components. It is particularly interesting to note that   the bright
M31 GC B514, which is the only other case we are aware of presenting an
extra-tidal component  falling as $r^{-3}$, also displays an obvious break in
the  profile near the King tidal radius \citep{lucky}. On the contrary, the SB
profile of NGC5694 appears perfectly  smooth and continuous over all its
declining branch, i.e from the core radius out to the last observed point. 
Finally, the density map shown in Fig.~\ref{map} does not show any obvious
elongation or characteristic S-shape structure that is typical of tidal tails
\citep[see, e.g.][]{mont}, and also the inner iso-density levels are remarkably
round  \citep[ellipticity $\epsilon=0.04$;][]{h96}.
An interesting analogy comes to mind with the case of the Galactic GC NGC1851, that \citet{ols} recently found to be surrounded by a low-SB halo extending out to
$\ga 250$~pc from the cluster center. Also in that case no sign of tidal tails
was found. However, at odds with the present case, in NGC1851 the onset of the extra tidal halo is marked by an obvious break in the profile, which, beyond the break, declines as $I\propto r^{-1.24}$.

It could be interesting, in this context, to estimate expected limiting radius imposed by the Galactic tidal field, or Jacobi radius ($r_t^J$), computed with the formula:

\begin{equation}
r_t^J = \frac{2}{3}\left(\frac{m_c}{2M_G}\right)^{\frac{1}{3}} R_{GC}
\end{equation}

that is formally correct for circular orbits within a logarithmic potential \citep[see][for references and discussion]{tid}.
$m_c$ is the cluster mass, derived from the total luminosity by adopting $\frac{M}{L_V}=2$, $R_{GC}$ is the galactocentric distance of the cluster, $M_G$ is the mass of the MW enclosed within $R_{GC}$, computed as $M_G=GV_{rot}^2R$. Assuming $V_{rot}=220$~km~s$^{-1}$, this gives a total mass of the MW in good agreement with the most recent analyses \citep{mwmass}. Adopting $R_{GC}=29.9$ we obtain 
$r_t^J= 194$~pc for NGC5694, more than a factor of 2 larger than the outermost point in our profile, at $r\simeq 93$~pc (assuming that the cluster do not extend much beyond this limit). Hence, at the present orbital phase the Galactic tidal field cannot have a serious impact on the structure of the cluster, at least in the region of the profile sampled by our data. The Jacobi radius would match  $r\simeq 93$~pc for a circular orbit at $R_{GC}\simeq 10$~kpc, thus suggesting that the peri-galactic point of the cluster orbit should be at a similar (or larger) distance, a quite plausible occurrence, given the large apo-galactic distance estimated by L06. 

McV interpreted the good overall performance of spherical and isotropic W75 models in fitting the profile of any kind of massive cluster even if extended, as an indication that {\em ``self-gravitating clusters commonly have envelope structures that do not match the extrapolations of simple K66 models fitting the cluster cores. This phenomenon is not confined exclusively to young clusters and is not obviously only transient; it may point instead to generic, internal cluster physics not captured by King's stellar distribution function''}. This can be the case also for NGC5694. An obvious example of physics that are not captured by K66 models is provided by the initial conditions at the cluster birth. These may leave their imprint even after a Hubble time in the outer fringes of the cluster, where the effects of two-body relaxation may bee too slow to completely erase them, and Galactic tides possibly have not played a major role. Further theoretical work is clearly needed to fully understand the nature of these features \citep{oh92,oh95}. 

In this framework NGC5694 appears as especially interesting, since the outer branch of its profile cannot be fitted even by W75 models, that are considered by McV as rather ad-hoc solutions.
In any case, the present result, as well as many other recent findings 
\citep[][for example]{chun,jg,pal14,ic4499}, suggest that there is still much to be learned about the outer structure of GCs, that can be revealed with modern wide field surveys allowing us to probe the very low SB regime of cluster profiles ($\mu_V\ga 26$~mag/arcsec$^2$), that is largely unexplored (see McV and NG06).

In conclusion we are unable to draw a firm
conclusion about the origin of the newly detected extended component
surrounding NGC~5694. It is clear, however, that the evidence presented here,
coupled with the results by L06, strongly indicate that this object deserves
further investigation, as it may fit into the scenario in which (at least some)
GCs are supposed to be the remnants of disrupted dwarf satellites, hence
one of the final end-states of the building blocks concurring to the
hierarchical  assembly of the MW \citep[see][and references
therein]{b08,boker,c10a}.  
Abundance analysis of a significant sample of NGC~5694 giants as well as a
photometric survey covering a significantly wider area than that explored here
are clearly needed to search for a metallicity spread and determine chemical
abundance ratios to shed more light on the nature of this very interesting, and
for a long time neglected, GC.

\section*{Acknowledgments}

M.B. and M.Co. acknowledge the financial support of INAF through the PRIN-INAF
2009 grant assigned to the project {\em Formation and evolution of massive star
clusters}, P.I.: R. Gratton. Support for M.Ca is provided by the Ministry of
Economy Development, and Tourism's Programa Inicativa Cient\'{i}fica Milenio
through grant P07-021-F, awarded to The Milky Way Millennium Nucleus; by
Proyecto Basal PFB-06/2007; by FONDAP Centro de Astrof\'{i}sica 15010003; and by
Proyecto FONDECYT Regular \#1071002.  
We thank the anonymous referee for its comments and suggestions.
We are grateful to A. Sollima for useful
discussions and suggestions, and for providing his code to compute Wilson models.  The Guide Star Catalogue-II is a joint project of
the Space Telescope Science Institute and the Osservatorio Astronomico di
Torino. Space Telescope Science Institute is operated by the Association of
Universities for Research in Astronomy, for the National Aeronautics and Space
Administration under contract NAS5-26555. The participation of the Osservatorio
Astronomico di Torino is supported by the Italian Council for Research in
Astronomy. Additional support is provided by European Southern Observatory,
Space Telescope European Coordinating Facility, the International GEMINI project
and the European Space Agency Astrophysics Division. This research has made use
of NASA's Astrophysics Data System.

\label{lastpage}


\begin{thebibliography}{999}
\bibitem[\protect\citeauthoryear{Bellazzini, Ferraro \& Ibata}{2003}]{bfi}
         Bellazzini, M., Ferraro, F.R., \& Ibata, R.A., 2003, AJ, 125, 188
\bibitem[\protect\citeauthoryear{Bellazzini}{2004}]{tid}
         Bellazzini, M., 2004, MNRAS, 347, 119	 
\bibitem[\protect\citeauthoryear{Bellazzini et al.}{2008}]{b08}
         Bellazzini, M., et al., 2008, AJ, 136, 1147	 
\bibitem[\protect\citeauthoryear{Belokurov et al.}{2006}]{bel5466}
         Belokurov, V., Evans, N.W., Irwin, M.J., Hewett, P.C., \& Wilkinson, M.I.,
	 2006, ApJ, 636, L29
\bibitem[\protect\citeauthoryear{B\"oker}{2008}]{boker}
         B\"oker, T., 2008m, ApJ, 672, L111	
\bibitem[\protect\citeauthoryear{Carretta et al.}{2010a}]{c10a}
         Carretta, E., et al., 2010, ApJ, 714, L7	  
\bibitem[\protect\citeauthoryear{Carretta et al.}{2010c}]{c10c}
         Carretta, E., et al., 2010, ApJ, 722, L1	 	  
\bibitem[\protect\citeauthoryear{Chun et al.}{2010}]{chun}
         Chun, S.H., et al., 2010, AJ, 139, 606	 
\bibitem[\protect\citeauthoryear{Cohen}{2004}]{cohen04} 
	Cohen, J.~G. 2004, AJ, 127, 1545
\bibitem[\protect\citeauthoryear{Cohen et al.}{2010}]{cohen}
         Cohen, J.G., Kirby, E.N., Simon, J.D., \& Geha, M., 2010, ApJ, 725, 288
\bibitem[\protect\citeauthoryear{Combes, Leon, \& Meylan}{1999}]{c99}
          Combes, F., Leon, S., \& Meylan, G., 1999, A\&A, 352, 149 
\bibitem[\protect\citeauthoryear{De Angeli et al.}{2005}]{dea}
         De Angeli, F., et al., 2005, AJ, 130, 116
\bibitem[\protect\citeauthoryear{Dotter et al.}{2010}]{dotter}
         Dotter, A.L., et al., 2010, ApJ, 708, 698
\bibitem[\protect\citeauthoryear{Elson, Fall \& Freeman}{1987}]{eff}
         Elson, R.A.W., Fall, S.M., \& Freeman, K.C., 1987, ApJ, 323, 54 (EFF)	 
\bibitem[\protect\citeauthoryear{Federici et al.}{2007}]{lucky}
         Federici, L., Bellazzini, M., Galleti, S., Fusi Pecci, F., Buzzoni, A.,
	 \& Parmeggiani, G., 2007, A\&A, 473, 429
\bibitem[\protect\citeauthoryear{Ferraro et al.}{1999}]{f99}
	 Ferraro, F.R., Messineo, M., Fusi Pecci, F., De Palo, A., Straniero,
	 O., Chieffi, A., \& Limongi, M., 1999, AJ, 118, 1738
\bibitem[\protect\citeauthoryear{Forbes \& Bridges}{2010}]{forbes}
         Forbes, D.A., \& Bridges, T., 2010, MNRAS, 404, 1203
\bibitem[\protect\citeauthoryear{Geisler et al.}{1995}]{geisler}
         Geisler, D., Piatti, A.E., Clari\'a, J.J., \& Minniti, D., 1995, AJ,
	 109, 605
\bibitem[\protect\citeauthoryear{Girardi et al.}{2005}]{trilegal}
         Girardi, L., Groenewegen, M.A.T., Hatziminaoglu, E., \& Da Costa, L.,
	 2005, A\&A, 463, 895
\bibitem[\protect\citeauthoryear{Gnedin \& Ostriker}{1997}]{gnedin}
         Gnedin, O.Y., \& Ostriker, J.P., 1997, ApJ, 474, 223	 	 
\bibitem[\protect\citeauthoryear{Grillmair et al.}{1995}]{grill95}
         Grillmair, C.J., et al., 1995, AJ, 109, 2553
\bibitem[\protect\citeauthoryear{Jordi \& Grebel}{2010}]{jg}
         Jordi, K., \& Grebel, E.K., 2010, A\&A, 522, A71	 
\bibitem[\protect\citeauthoryear{Johnston, Sigurdsson \& Hernquist}{1999}]{katy}
         Johnston, K.V., Sigurdsson, S., \& Hernquist, L., 1999, MNRAS, 302, 771
\bibitem[\protect\citeauthoryear{Harris \& Hesser}{1976}]{hh76}
         Harris, W.E., \& Hesser, J.E., 1976, PASP, 88, 377
\bibitem[\protect\citeauthoryear{Harris}{1996}]{h96}
         Harris, W.E., 1996, AJ, 112, 1487	 	 
\bibitem[\protect\citeauthoryear{Hazen}{1996}]{haz}
         Hazen, M.L., 1996, AJ, 111, 1184
\bibitem[\protect\citeauthoryear{King}{1966}]{k66}
         King, I.R., 1966, AJ, 71, 276 (K66)	 
\bibitem[\protect\citeauthoryear{Kirby et al.}{2011}]{kirby}
	 Kirby, E.~N., Cohen, J.~G., Smith, G.~H., Majewski, S.~R., Sohn, S.~T., 
	 \& Guhathakurta, P., 2011, ApJ, 727, 79 
\bibitem[\protect\citeauthoryear{Law \& Majewski}{2010}]{lm10}
         Law, D.R, \& Majewski, S.R., 2010, ApJ, 718, 1128
\bibitem[\protect\citeauthoryear{Lee, L\'opez-Morales \& Carney}{2006}]{l06}
         Lee, J.-W., L\'opez-Morales, M., \& Carney, B.W., 2006, ApJ, 646, L119
	 (L06)
\bibitem[\protect\citeauthoryear{Leon, Meylan \& Combes}{2000}]{leon}
         Leon, S., Meylan, G., \& Combes, F., 2000, A\&A, 359, 907
\bibitem[\protect\citeauthoryear{Mackey et al.}{2010}]{mackey}
         Mackey, A.D., et al, 2010, ApJ, 717, L11
\bibitem[\protect\citeauthoryear{Martin et al.}{2004}]{nic}
         Martin, N.F., Ibata, R.A., Bellazzini, M., Irwin, M.J., Lewis, G.F., \&
	 Dehnen, W., 2004, MNRAS, 248, 12
\bibitem[\protect\citeauthoryear{McLaughlin \& van der Marel}{2005}]{mcvdm}
         McLaughlin, D.E., \& van der Marel, R.P., 2005, ApJS, 161, 304
\bibitem[\protect\citeauthoryear{Milone et al.}{2008}]{milone}
         Milone, A., et al., 2008, ApJ, 673, 241
\bibitem[\protect\citeauthoryear{Monaco et al.}{2005}]{monaco05} 
	Monaco, L., et al., 2005, A\&A, 441, 141
\bibitem[\protect\citeauthoryear{Montuori et al.}{2007}]{mont}
         Montuori, M., Capuzzo-Dolcetta, R., Di Matteo, P., Lepinette, A., \&
	     Miocchi, P., 2007, ApJ, 659, 1212
\bibitem[\protect\citeauthoryear{Noyola \& Gebhardt}{2006}]{ng06}
         Noyola, E. \& Gebhardt, K., 2006, AJ, 132, 447 (NG06)
\bibitem[\protect\citeauthoryear{Odenkirchen et al.}{2001}]{pal5}
         Odenkirchen, M., et al., 2001, ApJ, 548, L165	 
\bibitem[\protect\citeauthoryear{Olszewski et al.}{2009}]{ols}
         Olszewski, E.W., Saha, A., Knezek, P., Subramanian, A., De Boer, T., \&
	 Seitzer, P., 2009, AJ, 138, 1570
\bibitem[\protect\citeauthoryear{Oh, Lin \& Aarseth}{1992}]{oh92}
         Oh, K.S., Lin, D.N.C., \& Aarseth, S.J., 1992, ApJ, 386, 506
\bibitem[\protect\citeauthoryear{Oh, Lin \& Aarseth}{1995}]{oh95}
         Oh, K.S., Lin, D.N.C., \& Aarseth, S.J., 1995, ApJ, 442, 142
\bibitem[\protect\citeauthoryear{Ortolani \& Gratton}{1990}]{og90}
         Ortolani, S., \& Gratton, R., 1990, A\&A Suppl., 82, 71 (OG90)
\bibitem[\protect\citeauthoryear{Perina et al.}{2009}]{perina}
         Perina, S., Federici, L., Bellazzini, M., Cacciari, C., Fusi Pecci, F.,
	 \& Galleti, S., 2009, A\&A, 507, 1375	 
\bibitem[\protect\citeauthoryear{Pritzl et al.}{2005}]{pritzl}
         Pritzl, B.~J., Venn, K.~A., \& Irwin, M., 2005, AJ, 130, 2140 
\bibitem[\protect\citeauthoryear{Robin et al.}{2003}]{r03}
         Robin, A.C., Reyl\'e, C., Derri\`ere, S., \& Picaud, S., 2003, A\&A, 
         409, 523 (R03)
\bibitem[\protect\citeauthoryear{Stetson}{1987}]{s87}
         Stetson, P.B., 1987, PASP, 99, 191
\bibitem[\protect\citeauthoryear{Stetson}{2000}]{s00}
         Stetson, P.B., 2000, PASP, 112, 926
\bibitem[\protect\citeauthoryear{Trager, King \& Djorgovski}{1995}]{tkd}
         Trager, S.C., King, I.R., \& Djorgovski, S., 1995, AJ, 109, 218 (TKD)
\bibitem[\protect\citeauthoryear{Sbordone et al.}{2007}]{sbord07}
	Sbordone, L., Bonifacio, P., Buonanno,R., Marconi, G., Monaco, L., 
	\& Zaggia, S., 2007, A\&A, 465, 815
\bibitem[\protect\citeauthoryear{Schlegel, Finkbeiner \& Davis}{1998}]{sfd}
         Schlegel, D.J., Finkbeiner, D.P., \& Davis, M., 1998, ApJ, 500, 525
\bibitem[\protect\citeauthoryear{Skrutskie et al.}{2006}]{2mass}
         Skrutskie, M.F., et al., 2006, AJ, 131, 1163
\bibitem[\protect\citeauthoryear{Sollima et al.}{2011}]{pal14}
         Sollima, A., Mart\'inez-Delgado, D., Valls-Gabaud, D., Pe\~narrubia,
	 J., 2011, ApJ, 726, 47
\bibitem[\protect\citeauthoryear{van de Ven et al.}{2006}]{ven}
	 van de Ven, G., van den Bosch, R.C.E., Vrolme, E.K., \& de Zeeuw, P.T.,
	 2006, A\&A, 445, 513	
\bibitem[\protect\citeauthoryear{Walker}{1994}]{m68}
	 Walker, A.R., 1994, AJ, 108, 555
\bibitem[\protect\citeauthoryear{Walker et al.}{2011}]{ic4499}
         Walker, A.R., et al., 2011, MNRAS, in press (arXiv:1103.4144)	 
\bibitem[\protect\citeauthoryear{Watkins et al.}{2010}]{mwmass}
         Watkins, L.L., Wyn Evans, N., \& An, J.H., 2010, MNRAS, 406, 246	 
\bibitem[\protect\citeauthoryear{Venn et al.}{2004}]{venn04}
	Venn, K.~A., Irwin, M., Shetrone, M.~D., Tout, C.~A., Hill, V.,
	\& Tolstoy, E., 2004, AJ, 128, 1177
\bibitem[\protect\citeauthoryear{Wilson}{1975}]{w75}
         Wilson, C.P., 1975, AJ, 80, 175 (W75)	 
\end{thebibliography}
\end{document}